\documentclass[sigconf]{acmart}
\AtBeginDocument{%
  }

\copyrightyear{2025}
\acmYear{2025}
\setcopyright{rightsretained}
\acmConference[CHI EA '25]{Extended Abstracts of the CHI Conference on Human Factors in Computing Systems}{April 26-May 1, 2025}{Yokohama, Japan}
\acmBooktitle{Extended Abstracts of the CHI Conference on Human Factors in Computing Systems (CHI EA '25), April 26-May 1, 2025, Yokohama, Japan}
\acmDOI{10.1145/3706599.3719984}
\acmISBN{979-8-4007-1395-8/2025/04}

\usepackage{acronym}
\usepackage{enumitem}

\begin{document}

\title[CART Through Correction of ASR]{Communication Access Real-Time Translation Through Collaborative Correction of Automatic Speech Recognition}

\author{Korbinian Kuhn}
\email{kuhnko@hdm-stuttgart.de}
\orcid{0009-0005-1296-4987}
\affiliation{%
  \institution{Stuttgart Media University}
  \city{Stuttgart}
  \country{Germany}
}

\author{Verena Kersken}
\email{kersken@hdm-stuttgart.de}
\orcid{0009-0000-5007-6327}
\affiliation{%
  \institution{Stuttgart Media University}
  \city{Stuttgart}
  \country{Germany}
}

\author{Gottfried Zimmermann}
\email{gzimmermann@acm.org}
\orcid{0000-0002-3129-1897}
\affiliation{%
  \institution{Stuttgart Media University}
  \city{Stuttgart}
  \country{Germany}
}

\renewcommand{\shortauthors}{Kuhn, Kersken, Zimmermann}

\begin{abstract}
Communication access real-time translation (CART) is an essential accessibility service for d/Deaf and hard of hearing (DHH) individuals, but the cost and scarcity of trained personnel limit its availability. While Automatic Speech Recognition (ASR) offers a cheap and scalable alternative, transcription errors can lead to serious accessibility issues. Real-time correction of ASR by non-professionals presents an under-explored CART workflow that addresses these limitations. We conducted a user study with 75 participants to evaluate the feasibility and efficiency of this workflow. Complementary, we held focus groups with 25 DHH individuals to identify acceptable accuracy levels and factors affecting the accessibility of real-time captioning. Results suggest that collaborative editing can improve transcription accuracy to the extent that DHH users rate it positively regarding understandability. Focus groups also showed that human effort to improve captioning is highly valued, supporting a semi-automated approach as an alternative to stand-alone ASR and traditional CART services.
\end{abstract}





\newacro{DHH}[DHH]{d/Deaf and hard of hearing}
\newacro{ASR}[ASR]{Automatic Speech Recognition}
\newacro{WER}[WER]{Word Error Rate}
\newacro{WPM}[wpm]{words per minute}
\newacro{NAD}[NAD]{National Association of the Deaf}
\newacro{E2E}[E2E]{end-to-end}
\newacro{RTLX}[RTLX]{NASA Raw Task Load Index}
\newacro{ML}[ML]{machine learning}
\newacro{CART}[CART]{Communication access real-time translation}
\newacro{SOTA}[SOTA]{state-of-the-art}

\maketitle

\section{Introduction}

Transcription plays a key role in providing \ac{DHH} individuals with access to spoken language. Real-time captioning presents significant challenges as text must be produced almost synchronously with speech. Professional services, known as \ac{CART}, typically rely on stenography or re-speaking (voice-writing). However, the scarcity and high cost of trained transcribers limit the availability of such services and, thus, the accessibility for many events. Advances in \ac{ASR} have greatly improved its accuracy, offering a scalable and cost-effective alternative \cite{Chorowski2015, Chan2015, Gulati2020, Peng2022, Radford2023}. As a result, automated captioning is now integrated into widely used applications and is increasingly relied upon as an accessibility service \cite{NDC2020}.

At the same time, \ac{DHH} individuals have consistently reported serious accessibility issues for automatically generated captions \cite{Deshpande2014, Kawas2016, Butler2019, NADPetition2019} and have criticised the lack of a declarative ruling regarding the quality of \ac{ASR} \cite{NADPetition2019, WCAG2024}. There seems to be a disparity between measured \ac{ASR} accuracy and \ac{DHH} individuals’ experience. One reason for this discrepancy is that benchmark scores can be misleading \cite{Geirhos2020}. The performance of an \ac{ASR} model in practice heavily depends on how well its training data cover the actual conditions \cite{Koenecke2020}. While high accuracy is achieved for General American English, noisy recording conditions \cite{Kinoshita2020}, specialised vocabulary \cite{Radford2023}, or a speaker’s accent \cite{Tadimeti2022} can significantly decrease the performance, particularly in lower-resourced languages \cite{Barrault2023}. Furthermore, commonly used evaluation metrics offer a limited representation of accessibility, leading to a mismatch between measured accuracy and the perceived transcription quality \cite{Wang2003, Mishra2011, Favre2013}. Until \ac{ASR} systems reliably produce accessible captions, \ac{CART} services involving humans are essential, and alternatives that address their limitations must be explored.

One promising yet under-explored approach to improving transcription accuracy is the manual correction of \ac{ASR} output \cite{Wald2006B, McDonnell2021, Seita2022, Kuhn2023}. As \ac{ASR} becomes more robust to noise and speaker variability, re-speaking benefits may diminish, allowing users to focus on corrections. This approach reduces the cognitive load on transcribers and could be performed by less trained personnel. This paper uses a mixed methods approach to evaluate a \ac{CART} workflow in which a group of non-professionals correct \ac{ASR} output in real-time to increase transcription accessibility. We developed a prototype that combines a live video stream, \ac{ASR} output, and collaborative text editing. A user study evaluates the feasibility and transcription accuracy of the approach. Complementary, \ac{DHH} focus groups were conducted to correlate their accuracy ratings with the user study results and to align their preferences with the capabilities of the workflow in general. Our results show:

\begin{itemize}
    \item Corrections can improve \ac{ASR} accuracy to the extent that \ac{DHH} readers rate it positively regarding understandability.
    \item Common evaluation metrics ignore quality aspects that focus groups consider equally important to word accuracy.
    \item The editing task is demanding and limits knowledge transfer.
\end{itemize}

\section{Background}

\subsection{Automatic Speech Recognition}

\ac{ASR} accuracy has improved significantly in recent years, with \ac{SOTA} models achieving high accuracy in many languages \cite{Radford2023, Barrault2023}. This progress has largely been driven by adopting \ac{E2E} models, particularly transformer architectures \cite{Vaswani2017, Chorowski2015, Chan2015, Gulati2020, Peng2022}, and machine learning techniques. Unsupervised \cite{Yu2020A}, semi-supervised \cite{Rasmus2015}, and self-supervised \cite{Schneider2019} learning approaches use web-scale training data to improve the robustness and overall quality of \ac{ASR} systems. Regarding accuracy rates, it has been suggested that \ac{ASR} has reached parity with manual transcription \cite{Xiong2017}. However, these results can be misleading, as optimising for specific datasets comes with the risk of shortcut learning \cite{Geirhos2020}. Despite improvements, the performance of \ac{ASR} can be limited by poor audio quality \cite{Kinoshita2020}, low-resource languages \cite{Radford2023, Barrault2023} and biased training data \cite{Koenecke2020, Tadimeti2022}. 

Studies indicate that the accuracy of automated captioning is insufficient to provide \ac{DHH} readers with full access to spoken content \cite{Deshpande2014, Kawas2016, Butler2019}. The \ac{NAD} has reported significant issues with captioning accuracy, timing, and completeness, and is seeking a declaratory ruling to address these concerns \cite{NADPetition2019}. Furthermore, the readability of ASR-generated captions can be challenging \cite{Butler2019, Liao2022}, especially when captions are not syntactically segmented \cite{Moron2018, Kushalnagar2018}. Streaming ASR further complicates the experience by producing intermediate results that cause distracting text changes in live captions \cite{Xingyu2023}.

\subsection{Manual Correction of ASR}

Correcting \ac{ASR} effectively reduces error rates \cite{Che2017} and has also been suggested by \ac{DHH} individuals \cite{McDonnell2021, Seita2022}. This method is already used by re-speakers, who use \ac{ASR} and manually correct the remaining errors. Correction is more efficient than creating a full manual transcription when the error rate is below 30\% \cite{Gaur2016}, a threshold that \ac{ASR} models can reliably achieve \cite{Addlesee2020, Radford2023}. While errors can be reduced by 30\% to 50\% \cite{Munteanu2008, Che2017, Than2021}, transcription quality can be higher as users tend to correct errors that have a greater impact on comprehension \cite{Kushalnagar2012, Bhavya2022}. \ac{DHH} readers also rated manually corrected captions highly accurate, sometimes preferring them over professional transcription services \cite{Deshpande2014}.

Real-time transcription is particularly challenging, as speech is up to ten times faster than typing \cite{Wald2006A}, and corrections must be made quickly \cite{Rander2010}. When working with non-professionals to overcome the shortage of professional transcribers \cite{Lasecki2012}, distributing the task among multiple individuals can reduce the demand and increase quality \cite{Harrington2013}. Studies indicate that at least three non-expert typists are required to correct \ac{ASR} transcriptions \cite{Takagi2015}, reducing errors by around 30\% \cite{Kuhn2023}. While real-time correction of \ac{ASR} has been proposed to increase accessibility \cite{Wald2006B, McDonnell2021, Seita2022}, further research is needed to assess its feasibility and effectiveness \cite{Kushalnagar2012, Deshpande2014, Kuhn2023}.

\subsection{Measuring Transcription Quality}

The accuracy of \ac{ASR} transcriptions is commonly measured by the \ac{WER}, which uses the Levenshtein distance on a word basis to determine the minimum edits between two texts \cite{Levenshtein1966}. The \ac{WER} is sensitive to non-semantic differences, treating them equally to missing or incorrect words. To address this, texts are often normalised by removing punctuation and converting to lowercase. Further steps, like unifying abbreviations, contractions, and written numbers, can reduce non-semantic errors \cite{Koenecke2020, Radford2023}. However, these modifications may remove aspects that affect comprehension, such as punctuation or semantic formatting.

The \ac{WER} has been criticised as a measure of accessibility, as it does not necessarily reflect human perception of accuracy \cite{Wang2003, Mishra2011, Favre2013}. Alternative metrics have been proposed but have not been widely adopted yet \cite{Morris2004, Apone2010, Kafle2019, Roux2022}. A key problem is that \ac{WER} does not consider the contribution of individual words to the overall understandability of a text \cite{Berke2018}. \Ac{ML}-based metrics address this issue by classifying the importance of words to determine the severity of errors \cite{Kafle2016, Amin2023}. While some studies report a higher correlation of such metrics with human perceptions of quality \cite{Kafle2017, Kafle2019}, adoption to other domains may be limited \cite{Wells2022}, and metrics may not necessarily perform better than the \ac{WER} \cite{Chavez2024}. Alternatives to automated metrics rely on manual classification of transcription errors \cite{Apone2010, Romero2015}. The usefulness of a transcript can also be assessed through subjective ratings, e.g. by ranking the understandability of a text. While qualitative measures are helpful for specific evaluations, they are difficult to obtain for large datasets and lack objectivity and replicability. 

In this study, we use the \ac{WER} as it is the standard metric in speech recognition research and to report accuracy with an objective measure. As it has received criticism as an accessibility metric, we additionally include accuracy ratings from a diverse group of \ac{DHH} individuals.

\section{Methodology}

We used a mixed methods approach to evaluate the proposed \ac{CART} workflow: Firstly, we conducted a user study in which non-professionals corrected \ac{ASR}-output in real-time. We measured the accuracy and tested different collaboration scenarios to gather information on efficacy, usability, and user experience. Secondly, we conducted focus groups with \ac{DHH} individuals to align their accuracy ratings with the results of the user study and their captioning requirements with the capabilities of the \ac{ASR} correction workflow.

\subsection{Recording}

The recording is an animated video which presents a scientific topic entertainingly.\footnote{https://www.youtube.com/watch?v=awwCj4NnGdk} We assume all participants have some knowledge about the general topic, but it is unlikely they are familiar with the specific content of the video. The recording lasts 10:14 minutes, and the reference transcript contains 1381 words, an average of 135 \ac{WPM}. We used a German video, as all participants were native speakers. The audio is a high-quality recording of a professional male speaker, with soft background music and sound effects. The speaker is not visible, and no lip reading is possible. The speech is grammatically correct, with clear pronunciation.

The video was transcribed with ten different \ac{ASR} models (see Appendix \ref{appendix:wer_by_chunk_and_service}). To increase the robustness of the \ac{WER}, we used a custom normaliser for German (see Appendix \ref{appendix:text_standardisation}), inspired by a related study \cite{Radford2023}. To start from a reasonable level of accuracy for the editing task, we selected an average transcript with a \ac{WER} of 9.3\%. The transcript was automatically formatted (see Appendix \ref{appendix:transcript_formatting}). The video was split into four evenly long parts for the \ac{DHH} groups’ caption rating. The parts contained captions with a \ac{WER} of 2\%, 5\%, 10\%, and 15\% (see Appendix \ref{appendix:wer_by_chunk_and_service}). The transcripts were then automatically formatted into captions (see Appendix \ref{appendix:caption_formatting}).

\subsection{User Study}

\subsubsection{Participants}

Overall, 75 people participated in the user study. Most were undergraduate computer science students, and the other participants were postgraduates and researchers at the same university. They gave informed consent to participate in the study and received a small thank-you gift. Participants were native German speakers and did not have a hearing disability. Participants worked in groups of three to edit the captions. 60 participants were randomly assigned to the different collaboration scenarios and editing roles by signing up for a specific time slot. An additional 15 participants were assigned to a control group.

\subsubsection{Collaboration Scenario}

Based on the results of previous research \cite{Piskorek2022, Kuhn2023}, we chose four collaboration scenarios to evaluate the impact of temporal and spatial task distribution on correction efficiency and editing conflicts. Temporal task distribution offsets each user's position in the transcript by delaying audio playback. Although latency must be kept to a minimum in real-time transcription, we wanted to evaluate the extent to which delayed corrections could improve accuracy. In some streaming scenarios, for example where users don’t interact with the presenters, near-live may be tolerable \cite{Moron2018}. Spatial task distribution mainly reduces user conflicts by assigning them different text segments. We set a group size of three, as suggested by related research \cite{Takagi2015}. For each scenario, we had five groups (i.e. 15 participants):

\begin{enumerate}[label={\Alph*}]
    \item Parallel: No task distribution mechanism has been applied.
    \item Delayed: Each user received the audio with a different delay (0, 10, 20 seconds).
    \item Chunked: Users were assigned chunks of the transcript and took turns correcting them.
    \item Mixed: Two users took turns correcting while the third user proofread with a 10-second audio delay.
\end{enumerate}

The scenarios were designed to help users distribute the work, but they did not prevent users from breaking out of their intended text sections. Editors could still help to correct other users' sections.

\subsubsection{Prototype}

\begin{figure}[t!]
  \centering
  \includegraphics[width=\linewidth]{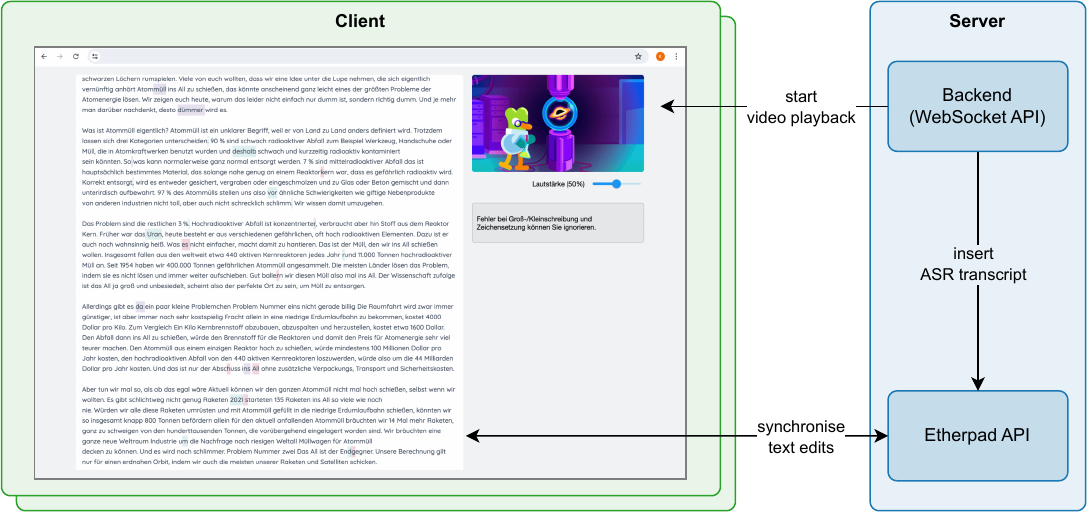}
  \caption{Architecture and user interface of the prototype}
  \Description{Architecture diagram with three components (Etherpad API, Backend API, Interface). The interface has a text input field on the left, and a video player on the right. The Backend API starts the interface’s video playback and inserts the ASR-generated transcript into the Etherpad API. The Etherpad API synchronises the text edits with the interface’s text input field.}
  \label{fig:architecture}
\end{figure}

The interface has an interactive text document on the left and a video player on the right (see Fig. \ref{fig:architecture}). In contrast to caption-based editing, our interface enables continuous text editing, allowing users to collaborate more freely \cite{Kuhn2023}. The design is functional and follows common caption editors and text editing solutions. An Etherpad is used to enable collaborative editing with real-time updates. The video player has only volume controls, and the server controls playback. The \ac{ASR} transcript is inserted consecutively and word by word, in sync with the audio track. Text insertions by the server are shown in the default black font colour. Text changes by users are highlighted in a user-specific colour to visually indicate the other participants’ edits.

\subsubsection{Procedure}

Participants scheduled a time for the user study in a computer lab or remotely via video conference. They used provided hardware or their own devices. Participants were informed about the study's goals, provided consent, and received a unique user ID for the editing software and questionnaire. All participants accessed the web application and completed a technical check to verify functionality. They were briefed on the collaboration scenario and instructed to focus on correcting word errors, ignoring capitalisation and punctuation. The video played at normal speed without interruption, and users edited the transcript without communicating with each other. After the video, participants completed a questionnaire including the \ac{RTLX} for cognitive load \cite{NASA1986, Grier2016}, questions on knowledge transfer from the video, and usability and user experience evaluations. They were then debriefed and given a small thank-you gift.

\subsubsection{Control Group}

To classify editing users’ cognitive load and knowledge transfer, a control group of 15 hearing participants watched the video without captions. After watching the video (without the option to rewind it), they answered a questionnaire with the \ac{RTLX} and the same questions regarding the videos’ content as the editing participants.

\subsection{DHH Focus Groups}

25 \ac{DHH} individuals using assistive technologies like hearing aids or cochlear implants participated in three focus groups. Two groups were local \ac{DHH} interest groups, one using spoken language and the other sign language, with diverse ages and occupations. The third group included university students from a \ac{DHH} initiative. Participants provided informed consent and received financial compensation. Despite diverse impairments, participants could be divided into two groups: those who relied solely on captioning and those who used captioning in addition to hearing. Focus groups were held in the groups' usual meeting places, with sign language interpreters assisting the sign language group.

Sessions began with a modified experience interview \cite{Zeiner2018}, where participants shared positive live captioning experiences using a handout for guidance. These experiences led to group discussions, which were transcribed and analysed for recurring themes and experiences \cite{Mayring2019}. Participants watched the same video used for the editing' study in four parts, each with captions at different accuracy levels. The video was shown on a projector with sound available in the room and via an induction loop. After each part, participants rated caption understandability on a seven-point Likert scale and gave written feedback. Final ratings and comments addressed accuracy, punctuation, formatting, latency, and verbatim transcription. Sessions concluded with debriefing.

\section{Results}

\subsection{Transcription Accuracy}

\begin{figure}[t!]
  \centering
  \includegraphics[width=\linewidth]{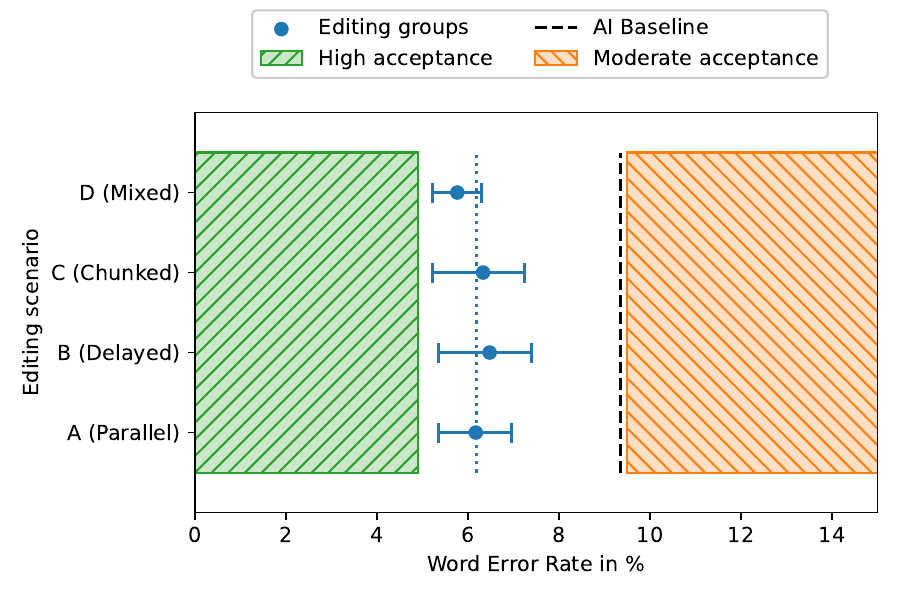}
  \caption{Word Error Rate by editing scenario}
  \Description{Graph showing the four editing scenarios in the user study on the y-axis, with the Word Error Rate from 0\% to 15\% on the x-axis. Each scenario is plotted as a range with upper and lower limits and the mean as a dot. The ASR baseline WER of 9.3\% and the scenarios average WER of 6.2\% are shown as vertical lines. Focus groups’ high acceptance ratings from 0\% to 5\% WER and moderate acceptance ratings from 9.5\% to 15\% WER are shown as rectangles. All editing scenarios have similar ranges of WER.}
  \label{fig:scenario_wer}
\end{figure}

Fig. \ref{fig:scenario_wer} shows the \ac{WER} of the \ac{ASR} transcript, the \ac{WER} after the corrections of each group, and the \ac{WER} ranges of the focus groups’ acceptance ratings. The initial \ac{WER} was 9.3\%. On average, the groups’ modifications resulted in a \ac{WER} of 6.2\% ($SD = 0.7\%$, $min = 5.2\%$, $max = 7.4\%$). A Kruskal-Wallis test was conducted to determine whether there is an effect of the editing scenario on the \ac{WER} after correction. There was no significant difference in the resulting \ac{WER} between the four different editing scenarios [$H(3) = 2.759$, $p = .430$] with a mean \ac{WER} score of 6.2\% for scenario A, 6.5\% for scenario B, 6.3\% for scenario C and 5.8\% for scenario D. On average, punctuation errors were reduced by 23\% ($SD=15\%$), and capitalisation errors by 15\% ($SD=16\%)$.

\subsection{Editing Users’ Comments}

Editing users’ comments on the questionnaire were categorised as positive or negative, and grouped into three topics: related to the general task, the usability of the prototype, and collaboration within the group. In total, 115 comments were negative, and 101 were positive. The number of comments grouped by topic was: task (negative = 56, positive = 28), usability (negative = 31, positive = 13), collaboration (negative = 28, positive = 60). 31 participants perceived the task as stressful or demanding, while 13 mentioned terms like “quite relaxing” or “rather easy”. 11 participants perceived correcting mistakes as rewarding. 37 participants commented positively on teamwork, 6 negatively, and 5 perceived teamwork as competitive. 16 participants mentioned editing conflicts with other team members, and 15 in following the continued video after correction.

\subsection{Cognitive Load}

The average \ac{RTLX} value for the editing users was 44.4 ($SD = 17.6$, $Mdn = 45.4$), and for the control group users was 13.4 ($SD = 11.2$, $Mdn = 12.5$). A Kruskal-Wallis test was conducted to determine whether there is an effect of the editing task on the perceived cognitive load. The results show a significant difference [$H(4) = 29.693$, $p < .001$]. A pairwise post-hoc Dunn-Bonferroni’s test found significant differences between participants in scenarios A, B, and C ($p < .001$) and D ($p = .023$), indicating that the editing groups perceived the task as significantly more stressful than the control group. No other differences were statistically significant.

\subsection{Knowledge Transfer}

On average, editing users answered 4 questions correctly ($SD = 1.5$, $Mdn = 4.0$) and control groups answered 5.5 questions correctly ($SD = 1.1$, $Mdn = 6.0$). A Mann-Whitney U test was conducted to determine whether there is a difference in the number of correct answers for editing participants and control group participants. There was a significant difference between the participant groups ($U = 202.5$, $p < .001$). Editing users rated their ability to follow the content on a seven-point Likert scale from 1 (very bad) to 7 (very good) on average with 3.6 ($SD = 1.5$, $Mdn = 3.0$). We analysed participants’ answers to how well they could follow the content while editing. 33 participants reported they could not follow the content at all, 14 participants could partly follow, and 9 could follow. 22 participants commented that they focused on hearing and reading to detect errors, not the actual content. 25 participants reported it was difficult or impossible to follow the content while correcting errors (or for a short time after a correction).

\subsection{Focus Groups' Ratings}

Participants rated the understandability of four video chunks with different \ac{WER} levels on a seven-point Likert scale from 1 (not understandable) to 7 (perfectly understandable). One participant did not fill out all scales and was excluded. A one-way ANOVA was conducted to determine the effect of the \ac{WER} on participants’ subjective rating of understandability. There was a significant effect of the video’s \ac{WER} on participants’ rating, indicating that participants rated a lower \ac{WER} more positively for understandability [$F(3, 92) = 8.710$, $p < .001$]. A Tukey’s HSD test revealed a significant difference between participants’ rating for video between a \ac{WER} of 4.9\% and 14.7\% ($p = .006$), between a \ac{WER} of 2.2\% and 14.7\% ($p < .001$), between a \ac{WER} of 4.9\% and 9.5\% ($p = .014$), and between a \ac{WER} of 2.2\% and 9.5\% ($p = .002$).

Participants rated the importance of different quality aspects of captions on a seven-point Likert scale from 1 (not important) to 7 (very important): accuracy ($M = 6.2$, $SD = 1.2$, $Mdn = 7.0$), punctuation ($M = 5.5$, $SD = 1.4$, $Mdn = 6.0$), capitalisation ($M = 5.9$, $SD = 1.6$, $Mdn = 7.0$), formatting ($M = 5.2$, $SD = 1.8$, $Mdn = 6.0$), latency ($M = 5.4$, $SD = 2.1$, $Mdn = 7.0$) and how verbatim captions have to be ($M = 5.6$, $SD = 2.2$, $Mdn = 7.0$).

\section{Discussion}

We tested an alternative \ac{CART} workflow in which non-professionals collaboratively correct \ac{ASR}. We also conducted focus groups with \ac{DHH} individuals to compare their preferences for live captioning with the semi-automated transcription process. We measured caption quality using the \ac{WER} and with ratings from \ac{DHH} caption readers, as there is currently no objective measure for the accessibility of captions.

All editing groups could reduce the \ac{WER} of the \ac{ASR} transcript. The initial \ac{WER} was 9.3\%, a level that reached medium acceptance by the \ac{DHH} focus groups. The corrected transcripts achieved an average \ac{WER} of 6.2\%, similar to the accuracy range rated positively regarding understandability. The relative error reduction of 33\% is consistent with findings from related studies \cite{Munteanu2008, Che2017, Than2021, Kuhn2023}. Although users were instructed to focus on word errors, punctuation and capitalisation errors also decreased, suggesting that people tend to correct more meaningful errors, which are not measured by \ac{WER} but enhance comprehension (cf. \cite{Kushalnagar2012, Bhavya2022}). The comprehensibility of the transcript may therefore be higher than the \ac{WER} suggests.

The collaboration scenario had no significant effect on the resulting accuracy of the transcript or the perceived cognitive load of the users. While teamwork was perceived as stress-reducing, a mechanism to guide collaboration seems necessary, as most users within the parallel scenario reported editing conflicts. Editors perceived working in chunks as more positive, as it reduced such conflicts and allowed them more time to review their segment of the text. Since working with delays did not improve performance, and the \ac{DHH} focus groups emphasised their preference for minimal latency (cf. \cite{McDonnell2021}), temporal task distribution seems to be neither an effective nor a viable solution. Correcting mistakes was perceived as rewarding, but participants also mentioned a sense of competition. A revised interface could emphasise the task's collective goal, for example, by displaying the total number of corrections made.

The editor’s average \ac{RTLX} score was 44, which is not overly demanding but does require some effort \cite{Grier2016}. The score was significantly higher than that of the control group, who only watched the video. In the quiz following the video, editors gave significantly fewer correct answers than participants in the control group. Most editors reported that they could not or only partially follow the content, especially whilst or shortly after correcting errors. This suggests that following the content of the video and correcting errors in the script is too demanding to be done simultaneously \cite{Kushalnagar2014}. Therefore, the task of editing should be performed by dedicated editors so as not to disadvantage other listeners \cite{McDonnell2021}.

Reading error-free captions is already a demanding task because the speed of speech often exceeds that of reading. Any mistakes, such as word errors, missing punctuation, incorrect capitalisation, or poor formatting, disrupt the reading flow and require mental compensation that can only be managed to a limited extent \cite{Nam2020}. Acceptance ratings among \ac{DHH} users varied significantly depending on the \ac{WER}; captions with 2-5\% \ac{WER} were rated significantly higher than those with 10-15\% \ac{WER} (cf. \cite{Butler2019}). Positive comments were more frequent for captions with fewer errors, while higher error rates led to more negative feedback. The impact of errors on reading speed is also reflected in the \ac{DHH} participants' negative comments on talking speed in video segments with higher \ac{WER}, despite similar \ac{WPM} rates across segments.

\ac{DHH} individuals rated accuracy, capitalisation, punctuation, and semantic formatting as equally important for understandability (cf. \cite{Butler2019, Kushalnagar2018}), supporting the criticism of \ac{WER} as a limited measure of accessibility \cite{Wang2003, Mishra2011, Favre2013, Chavez2024}. The importance of different quality aspects is strengthened by the fact that captions with lower \ac{WER} also had fewer punctuation and capitalisation errors. Minimal latency was strongly preferred, as caption reading is often combined with lip-reading or hearing (cf. \cite{McDonnell2021}). Focus groups also expressed a clear preference for verbatim transcription. While \ac{ASR} systems can provide low-latency verbatim output, corrections remain essential to address errors that hinder comprehension. Combining the speed of \ac{ASR} with human editing is a promising alternative to traditional \ac{CART} workflows to improve accessibility until fully automated systems can reliably meet user needs.

Participants expressed predominantly negative experiences with \ac{ASR}-generated captions and their preference for professional caption services: \textit{"I can clearly see whether it is people who make an effort to write everything down or whether only ASR is used"}. \ac{ASR} was seen as frustrating and unreliable, described as \textit{"very irritating"} and producing "wondrous linguistic creations". In contrast, human involvement was highly valued. Participants stated, \textit{"A human being should still stand between AI and us"}, and found their transcription errors more tolerable: \textit{"Mistakes made by written interpreters are forgivable"}. The discussions showed that automated captioning lacks accessibility, while human effort is highly valued. An \ac{ASR} correction workflow could bridge this gap and provide a more acceptable solution for \ac{DHH} readers.

\section{Conclusion}

\ac{ASR} offers a great opportunity to improve the accessibility of events through real-time captioning, but its lack of accuracy limits the participation of people who rely on it. Collaboratively correcting \ac{ASR} could improve accessibility, especially when professional transcribers are unavailable. We evaluated this alternative \ac{CART} workflow in a user study and conducted focus groups to align its capabilities with the preferences of \ac{DHH} individuals.

The user study showed that the initial \ac{WER} of \ac{ASR} can be reduced to the range of high acceptability, even though corrections must be made in real-time. As humans tend to correct errors that influence the meaning of a text, the accessibility may be even higher than the \ac{WER} indicates. As real-time correction of \ac{ASR} is demanding, it requires dedicated personnel apart from the audience to ensure that knowledge transfer is not hindered. Correcting as a group can make the task more manageable, and spatial task distribution can reduce stress and editing conflicts. While relying solely on \ac{ASR} may feel like a second-rate accommodation for \ac{DHH} individuals, human efforts to improve their participation were highly valued, potentially increasing the acceptance of \ac{ASR} in a semi-automated captioning workflow.

\section{Limitations \& Future Work}

Our findings are based on a single video presentation with a professional speaker, limiting generalisability to other contexts. Less favourable conditions impact \ac{ASR} accuracy, complicate manual correction, and reduce intelligibility for caption readers. \ac{DHH} participants did not rate the edited captions directly to avoid bias from individual editing groups. At this stage of the research, we believe that comparing the \ac{WER} after editing with the \ac{WER} ratings of \ac{DHH} individuals is a more objective and transferable approach to other studies. Future research should evaluate the workflow end-to-end, with editors correcting streaming \ac{ASR} and \ac{DHH} users reading and rating the corrected transcript. It could also be explored how AI models can assist humans by providing autocompletions, suggestions or autonomous corrections.

\clearpage

\begin{acks}
We thank all participants for their time and contribution to this research. We are especially grateful to everyone participating in the focus groups and their organisations (Schwerhörigenverein Stuttgart e.V.; DHH students from PH Heidelberg; Barrierefreie Kommunikation e. V.). This research was conducted as part of the SHUFFLE Project (“Hochschulinitiative Barrierefreiheit für alle”) and funded by “Stiftung Innovation in der Hochschullehre”.
\end{acks}

\bibliographystyle{ACM-Reference-Format}
\bibliography{bibliography}

\appendix

\section{ASR Transcriptions}
\label{appendix:wer_by_chunk_and_service}

Table \ref{tab:wer_by_chunk_and_service} displays the \ac{WER} with text normalisation for each video chunk and the whole transcript. Bold marked entries were used for the user study (overall) and the \ac{DHH} readers’ caption rating (per chunk).

\setlength{\tabcolsep}{3pt}
\begin{table}[ht!]
  \caption{Word Error Rate in percent by ASR model}
  \label{tab:wer_by_chunk_and_service}
  \begin{tabular}{lrrrrr}
    \toprule
    Service & Overall & Chunk 1 & Chunk 2 & Chunk 3 & Chunk 4 \\
    \midrule
    Amazon & 8.62 & 7.06 & 9.70 & \textbf{9.46} & 8.10 \\
    AssemblyAI & 8.91 & 10.59 & 11.59 & 7.45 & 5.61 \\
    Deepgram & 16.00 & 15.88 & 17.25 & 13.18 & 17.76 \\
    Google & 12.96 & \textbf{14.71} & 12.13 & 10.32 & 14.95 \\
    IBM & 16.00 & 17.35 & 15.63 & 11.17 & 20.25 \\
    Microsoft & 6.37 & 7.06 & 8.63 & 5.44 & 4.05 \\
    Rev AI & \textbf{9.34} & 9.12 & 9.97 & 9.46 & 8.72 \\
    Speechmatics & 3.69 & 2.65 & \textbf{4.85} & 4.30 & 2.18 \\
    Speechtext.AI & 11.08 & 10.88 & 15.36 & 8.60 & 9.03 \\
    Whisper & 4.06 & 7.94 & 2.70 & 3.44 & \textbf{2.18} \\
    \midrule
    Average & 9.70 & 10.32 & 10.78 & 8.28 & 9.28 \\
    \bottomrule
  \end{tabular}
\end{table}
\setlength{\tabcolsep}{6pt}

\section{Text Standardisation}
\label{appendix:text_standardisation}

The transcripts were automatically normalised before calculating the WER according to the following rules: 

\begin{enumerate}
    \item Replace all white space characters with a space and remove repeated spaces.
    \item Replace German diacritics (e.g. ä -> ae, ß -> ss).
    \item Convert currencies and symbols to their written forms (e.g. € -> Euro, \% -> Prozent).
    \item Replace common German abbreviations (e.g. z.B. -> zum Beispiel).
    \item Convert common German contractions to its long form (e.g. gibt’s -> gibt es).
    \item Convert all Arabic numbers to a numeric expression (e.g. 200 -> zweihundert).
    \item Transform text to lowercase and remove all punctuation.
    \item Remove repeated, leading, and trailing spaces.
\end{enumerate}

\section{Transcript Formatting}
\label{appendix:transcript_formatting}
The transcript was automatically formatted according to the following rules:

\begin{itemize}
    \item Start a new paragraph after a silence of 5 seconds.
    \item 800 to 1000 characters per paragraph (100 to 200 for chunked scenarios).
    \item End paragraph on sentence ending characters (.!?;).
\end{itemize}

\section{Caption Formatting}
\label{appendix:caption_formatting}
The captions were automatically formatted according to the following rules:

\begin{itemize}
    \item Length of 60 to 80 characters.
    \item End caption if a sentence ending character appears (.!?;).
    \item Format the caption into two lines:
    \begin{itemize}
        \item Break in between the character ranges of 30 to 60.
        \item Try to break at a punctuation character (.,!?;:).
        \item Otherwise, break at any non-word character (e.g. white space).
    \end{itemize}
\end{itemize}

\end{document}